\documentclass{article}
\pdfoutput=1

\usepackage[margin=1in]{geometry}
\usepackage{amsmath, amssymb, amsthm, amsfonts, mathtools}
\usepackage{microtype}
\usepackage{authblk}
\usepackage{xcolor}
\usepackage[colorlinks=true,linkcolor=blue,urlcolor=blue,citecolor=blue]{hyperref}

\theoremstyle{plain}

\theoremstyle{definition}

\numberwithin{equation}{section}

\newcommand{\C}{\mathbb{C}}
\newcommand{\Z}{\mathbb{Z}}
\newcommand{\HH}{\mathbb{H}}
\newcommand{\Sp}{\operatorname{Sp}}
\newcommand{\Imag}{\operatorname{Im}}
\newcommand{\code}[1]{\texttt{#1}}

\usepackage[T1]{fontenc}
\usepackage{lmodern}
\usepackage{fancyvrb}
\fvset{fontsize=\footnotesize,xleftmargin=0.5em}

\DeclareUnicodeCharacter{2102}{\ensuremath{\mathbb{C}}}   
\DeclareUnicodeCharacter{2115}{\ensuremath{\mathbb{N}}}   
\DeclareUnicodeCharacter{211D}{\ensuremath{\mathbb{R}}}   
\DeclareUnicodeCharacter{2124}{\ensuremath{\mathbb{Z}}}   
\DeclareUnicodeCharacter{210D}{\ensuremath{\mathbb{H}}}   
\DeclareUnicodeCharacter{0391}{\ensuremath{A}}            
\DeclareUnicodeCharacter{0393}{\ensuremath{\Gamma}}       
\DeclareUnicodeCharacter{0394}{\ensuremath{\Delta}}       
\DeclareUnicodeCharacter{039B}{\ensuremath{\Lambda}}      
\DeclareUnicodeCharacter{03A3}{\ensuremath{\Sigma}}       
\DeclareUnicodeCharacter{03B1}{\ensuremath{\alpha}}       
\DeclareUnicodeCharacter{03B2}{\ensuremath{\beta}}        
\DeclareUnicodeCharacter{03B3}{\ensuremath{\gamma}}       
\DeclareUnicodeCharacter{03B4}{\ensuremath{\delta}}       
\DeclareUnicodeCharacter{03BB}{\ensuremath{\lambda}}      
\DeclareUnicodeCharacter{03C4}{\ensuremath{\tau}}         
\DeclareUnicodeCharacter{03C9}{\ensuremath{\omega}}       
\DeclareUnicodeCharacter{2200}{\ensuremath{\forall}}      
\DeclareUnicodeCharacter{2203}{\ensuremath{\exists}}      
\DeclareUnicodeCharacter{2227}{\ensuremath{\wedge}}       
\DeclareUnicodeCharacter{2228}{\ensuremath{\vee}}         
\DeclareUnicodeCharacter{00AC}{\ensuremath{\neg}}         
\DeclareUnicodeCharacter{2208}{\ensuremath{\in}}          
\DeclareUnicodeCharacter{2209}{\ensuremath{\notin}}       
\DeclareUnicodeCharacter{2229}{\ensuremath{\cap}}         
\DeclareUnicodeCharacter{222A}{\ensuremath{\cup}}         
\DeclareUnicodeCharacter{2286}{\ensuremath{\subseteq}}    
\DeclareUnicodeCharacter{2260}{\ensuremath{\ne}}          
\DeclareUnicodeCharacter{2264}{\ensuremath{\le}}          
\DeclareUnicodeCharacter{2265}{\ensuremath{\ge}}          
\DeclareUnicodeCharacter{2243}{\ensuremath{\simeq}}       
\DeclareUnicodeCharacter{2245}{\ensuremath{\cong}}        
\DeclareUnicodeCharacter{2223}{\ensuremath{\mid}}         
\DeclareUnicodeCharacter{2192}{\ensuremath{\to}}          
\DeclareUnicodeCharacter{2194}{\ensuremath{\leftrightarrow}} 
\DeclareUnicodeCharacter{21D1}{\ensuremath{\Uparrow}}     
\DeclareUnicodeCharacter{21D2}{\ensuremath{\Rightarrow}}  
\DeclareUnicodeCharacter{2202}{\ensuremath{\partial}}     
\DeclareUnicodeCharacter{2211}{\ensuremath{\sum}}         
\DeclareUnicodeCharacter{2218}{\ensuremath{\circ}}        
\DeclareUnicodeCharacter{2022}{\ensuremath{\bullet}}      
\DeclareUnicodeCharacter{2212}{\ensuremath{-}}            
\DeclareUnicodeCharacter{2080}{\ensuremath{{}_0}}         
\DeclareUnicodeCharacter{2081}{\ensuremath{{}_1}}         
\DeclareUnicodeCharacter{2082}{\ensuremath{{}_2}}         
\DeclareUnicodeCharacter{2097}{\ensuremath{{}_l}}         
\DeclareUnicodeCharacter{00B2}{\ensuremath{{}^2}}         
\DeclareUnicodeCharacter{1D9C}{\ensuremath{{}^{c}}}       
\DeclareUnicodeCharacter{1DA0}{\ensuremath{{}^{f}}}       
\DeclareUnicodeCharacter{2014}{---}                       
\DeclareUnicodeCharacter{03B8}{\ensuremath{\theta}}       
\DeclareUnicodeCharacter{03B6}{\ensuremath{\zeta}}        

\title{Axioms for physical reasoning:\\ codifying the Seiberg--Witten solution in Lean%
  \thanks{Working paper, draft. Companion to the formal development at
  \href{https://github.com/mrdouglasny/seiberg-witten}{\code{mrdouglasny/seiberg-witten}}
  and to the ICML 2026 talk \emph{Validation of AI-generated results in theoretical physics}.}}
\author[1]{Michael R. Douglas \thanks{mdouglas@cmsa.fas.harvard.edu}}
\affil[1]{Center Of Mathematical Sciences And Applications, Harvard University, Cambridge, MA 02138 USA}
\date{July 2026}

\begin{document}
\maketitle

\begin{abstract}
Mathematicians have embraced interactive theorem provers with growing enthusiasm---building large
shared libraries and machine-checking a string of landmark results. Theoretical physics is
different: most of its results are not theorems but justified by arguments the community trusts
without a rigorous proof. For many---the one we treat here among them---no rigorous
proof is within reach. For 4d Yang--Mills theory, deriving exact rigorous results from first principles
would first require constructing the interacting theory nonperturbatively, which is a sizable piece of
one of the Clay Millennium prize problems. 

We argue here that an interactive theorem prover can be used to verify some non-rigorous physics arguments.
The method is to postulate a short list of \textbf{explicit, named physical
postulates}, which imply the physical results by virtue of a machine-checkable proof.
The trust that remains then rests on that short, inspectable list, and the prover can
report, for any downstream result, exactly which assumptions it used. We carry this out for
the Seiberg--Witten solution of $\mathcal{N}=2$ $SU(2)$ super-Yang--Mills---the genus-one
case---formalized in Lean 4; the higher-genus $SU(N)$ generalization is developed in the same
repository as an axiomatized skeleton and left to future work. We describe what is proved, what is assumed, how the assumptions are checked---external
review and an independent numerical oracle---and why this discipline is a sound standard for
validating AI-generated results in theoretical physics. What we offer is a \emph{discipline},
reviewable on its own terms: a reader may take the Seiberg--Witten mathematics on trust and
still assess the formalization method.
\end{abstract}

\tableofcontents

\section{Introduction}

Most of what theoretical physics asserts is not rigorously proved. Across the field, the results we build on
are typically not theorems but conclusions of \emph{arguments}---chains of physical reasoning
trusted because they cohere internally, because independent routes to the same quantity agree, and
ultimately because they account for experiment. A result can be as certain as anything in science
and still have no rigorous proof from first principles; certainty in physics rests on a different footing
than in mathematics.

In quantum field theory the gap is very wide. There the deepest results---the spectrum of a
gauge theory, its phase structure, an exact low-energy effective action---follow from reasoning
based on principles like symmetry, locality, dualities and consistency. The {Seiberg--Witten
solution}~\cite{SeibergWitten1994a,SeibergWitten1994b}, our
subject below, is a paradigmatic example.  A rigorous proof is out of reach: it would require first
constructing the interacting theory nonperturbatively, on the scale of the Yang--Mills
Millennium Prize. Such results are as certain as anything in physics, yet they are not theorems.

This gap has become pressing for a new reason: AI systems now generate physical derivations, and
their characteristic error is not a wrong algebra step---these can be caught automatically---but a
\emph{hallucinated assumption}, an unphysical premise or a dropped subtlety presented as if derived.
The possibility of such errors leads many to insist that AI-generated proofs require human auditing, but 
reading the entire proof is often not practical. 
What one wants instead is a way to make an
argument's assumptions explicit and machine-auditable, cleanly separating what is genuinely derived
from what is taken on faith.

Interactive theorem provers are built for exactly that separation. A technology from computer
science---which grew out of hardware and software verification---proof assistants have lately
drawn intense interest in mathematics, with fast-growing libraries and a number of landmark theorems now machine-checked.
Introductions to and reviews of the use of interactive theorem proving in mathematics
include~\cite{Buzzard2022,Avigad2024,Tao2025}, and in physics
include~\cite{ToobySmith2025,ToobySmith2025Perspective,douglas_formalization_2026};
a recent formalization of physics reasoning in Lean---adjacent in spirit to the present
work---poses model-building questions as formal statements~\cite{KrippendorfToobySmith2026}.
We will demonstrate here how to apply these tools to a physics argument. A proof assistant such as \textbf{Lean 4} is a
language in which one writes definitions and proofs that a small, fixed {kernel}---a few
thousand lines of code, audited once and then trusted---checks line by line. A statement the kernel accepts is correct \emph{relative to its
inputs}: the handful of logical axioms built in,\footnote{Examples are the axiom of choice, and excluded middle:
for any proposition $P$, either $P$ or $\mbox{not}\,P$ must be true.
} 
plus any further {axioms} the author explicitly declares. 

Throughout this paper, ``axiom'' is meant in the logician's sense --- a \emph{formalized, named, audited
assumption} the kernel tracks.\footnote{  In Lean this can be done in several ways; as an
explicit \code{axiom} stating the assumption; by using the \code{sorry} tactic which accepts an intermediate result
without proof; or as hypotheses which are not assumed globally but on which the results of interest are proven conditionally.
In this introduction we use ``axiom'' to refer to any of these methods.}
Our work here will use assumptions of two kinds, and we name them differently:
\textbf{mathematical axioms} --- true classical mathematics assumed without formal proof --- and
\textbf{physical postulates}, the formal translations of physical assumptions. (The title's
``axioms'' is the umbrella, logician's sense just described; within it, these two names
distinguish what kind of trust each assumption asks for.) The distinction is
enforced by the architecture, not just the prose: in the development the Lean \code{axiom}
keyword carries only mathematics, while the physical postulates are predicates carried in the
theorems' types.
To explain the first, in formalizing a mathematical proof, one sometimes provisionally assumes provable intermediate steps as axioms, in order
to check that the final theorem is provable before going back to complete these intermediate steps.  For example,
in our OSforGFF project \cite{douglas_formalization_2026} we assumed Minlos' theorem in our first release, and proved it in the second.
This leads to practical issues but, given that we only axiomatize provable correct results, no new issues of principle.

The physical postulates are formal translations of physical assumptions.   We will explain the ones we use here in detail below,
but they include the basic definitions of physical concepts in mathematical terms and their accepted properties.  These might
be provable from more basic physical assumptions; they might encode empirical data; they might be believed because they
are true in many previously studied cases; or they might be pure leaps of intuition.  The point here is that
by stating these postulates formally,
they become {precise, finite, named, and dependency-tracked}. 

The combination of mathematical axioms and physical postulates 
assumed in deriving a result is the {trusted base}---everything one must believe in
order to believe the result---and the Lean prover can print it: for any theorem, the command
\code{\#print axioms} reports exactly which axioms it used, and nothing reaches that list without
being named. Modern provers ship with large formal libraries of mathematics---Lean's is
\textbf{Mathlib}~\cite{mathlib}, several hundred thousand theorems---so one formalizes atop real
analysis, algebra and geometry rather than from scratch.

The point of this paper is to turn that machinery on a physics result that \emph{cannot} be proved,
and to use it honestly. We do not try to rigorously prove the Seiberg--Witten solution; that is beyond reach.
Instead we \textbf{state the physical reasoning as a short list of explicit, named postulates, prove
every mathematical consequence of them in Lean, and let the kernel verify the relation between the two.}
The residual trust is then concentrated on that named list.  
The \code{\#print axioms} command audits, for any downstream claim, precisely which
physical inputs it rests on. 
In principle, the kernel could expose contradictions
hidden among the axioms (a logical contradiction implies all propositions including manifestly false ones).
Justifying the axioms is still the physicist's task---here, external
review and an independent numerical oracle---but it becomes a finite, inspectable one rather than a
diffuse act of faith.

We chose the {Seiberg--Witten solution} as a test case for many
reasons.  It is a
celebrated, universally trusted result, unprovable by present-day mathematics, yet mathematically precise
in content---it identifies the exact low-energy physics with the period geometry of a specific family
of algebraic curves, so both the physical assumptions and their mathematical consequences are crisp
enough to formalize. Around it we obtain a clean split: a proved mathematical spine (the curve, its
genus, positivity of the special-K\"ahler metric, the explicit genus-one solution with its
special coordinates and one-loop running), the physical hypotheses as named predicates, and the
headline statements---existence, and uniqueness up to duality---proved from those hypotheses
plus classical, citable mathematical inputs. We treat the genus-one ($SU(2)$) case in full;
the higher-genus family is future work (\S6).

The rest of the paper proceeds as follows. \S2 sets out the two kinds of formalization a physics
result calls for and the bookkeeping discipline they impose. \S3 states the Seiberg--Witten solution
compactly. \S4 is the formalization proper---the math$\leftrightarrow$physics dictionary, what is assumed (the
hypotheses, with their physical origins), what is proved, what formalizing forced into the open, and a worked axiom-free check
(Argyres--Douglas points in a worked matter example).
\S5 describes how we check the axioms. \S6 lays out the remaining mathematical debt and our
posture toward it---the genus-one classical axioms we deliberately keep, and the higher-genus
programme we leave to future work---and \S7 the prospects for deriving the physical assumptions
themselves from deeper principles.
\S8 concludes. Two appendices give the Lean statements of the physical
hypotheses (A) and the residual mathematical axioms (B).

\section{Two kinds of formalization}

A mathematician who formalizes a theorem has a proof to formalize; a physicist who formalizes a
\emph{result} usually has none. The two situations call for two different kinds of formalization:

\begin{center}
\renewcommand{\arraystretch}{1.3}
\begin{tabular}{@{}p{0.26\textwidth}p{0.30\textwidth}p{0.34\textwidth}@{}}
\hline
& \textbf{What it is} & \textbf{Examples} \\
\hline
\textbf{Mathematical physics} & A rigorous statement exists; you prove it. & Stability of
matter; constructive QFT; the lattice-gauge mass gap---\emph{and the proved spine of this
project}: the curve, its genus, special-K\"ahler positivity, the period matrix. \\
\textbf{Foundations for physical reasoning} & Codifying the \emph{non-rigorous} reasoning the
community trusts, where no proof is within reach. &
\emph{The Seiberg--Witten argument itself.} \\
\hline
\end{tabular}
\end{center}

Mathematical physics formulates and proves physical results within the accepted foundations of mathematics,
so the first kind of formalization can follow existing practice for formalizing mathematics.
But Seiberg--Witten requires the second kind: to date, there is no rigorous derivation that the low-energy
physics on the Coulomb branch of $\mathcal{N}=2$ $SU(N)$ gauge theory is governed by the curve
$y^2 = P_N(x)^2 - \Lambda^{2N}$~\cite{KLYT1995,ArgyresFaraggi1995}. In its place stands an \emph{argument}---holomorphy from
supersymmetry, the exact pattern of singularities, electric--magnetic duality, the weak-coupling
asymptotics---that together single out that curve.  This result was confirmed by independent arguments, most notably
Nekrasov's instanton computation of the prepotential \cite{Nekrasov2003}.  This calculation and
the proof that it reproduces the original Seiberg-Witten solution has been made rigorous
\cite{NekrasovOkounkov2006,NakajimaYoshioka2005,BravermanEtingof2006}; however the claim that it actually
describes $N=2$  super Yang-Mills rests on the physical assumption that the instanton expansion produces an exact result,
whose rigorous statement and proof would require some rigorous definition of $N=2$  super Yang-Mills,
so is out of reach.

We call the practices and axioms used in this second category of formalization
\textbf{foundations for physical reasoning}, by analogy with
foundations of mathematics.
This distinction is not a value judgment; as we have discussed at length elsewhere \cite{douglas2011foundations}, both
non-rigorous and rigorous arguments have their advantages and disadvantages.  The paradigm we propose here could be considered
a hybrid approach, making as much of the physics argument as possible rigorous, and making explicit all the ingredients which go beyond
mathematical rigor as it exists now.

To realize this explicitness, a certain discipline should be followed.  No physical input should hide in a mathematical definition,
so {no mathematical axiom may use physical vocabulary}. Since not all of the mathematics needed for Seiberg-Witten has been formalized,
we postulate mathematical axioms, but these name only mathematics
(periods, the period matrix, monodromy in $\Sp(2g,\Z)$, asymptotics).
Every physical quantity---effective coupling, central charge, electric--magnetic duality, beta function---will have an
\textbf{explicit definition} identifying it with a mathematical object, collected in one dictionary
and never silently equated. The trusted mathematical base is then logic plus \emph{purely mathematical} axioms;
physics enters only through inspectable definitions and the named hypotheses H0--H7. 

An example is the beta function for the effective
coupling.  The mathematical input---the logarithmic weak-coupling running of the period
ratio---is a \emph{theorem}: for the explicit genus-one coupling,
$u\,d\tau/du \to i/\pi$ is proved, a corollary of the (also proved) Legendre relation.  The
identifications \code{coupling} $:=$ period ratio and
\code{betaFunction} $:= \Lambda\, d/d\Lambda\, \tau$ are definitions, and the one-loop value
$b_0 = 2N - N_f$ follows as a theorem of the three logical axioms. The consolidated period axiom of Appendix~B is in this form too: it is a
purely \emph{mathematical} statement---the existence and rigidity of the period geometry of the curve
family $y^2 = P^2 - \Lambda^{2N}$ (period charts, polarized period data, monodromy in $\Sp$,
prescribed asymptotics)---whose physical reading, that this data \emph{is} the SW effective theory
with electric--magnetic duality the symplectic reframing, is supplied entirely by the dictionary
(\S4.1). What is \emph{not} done is \textbf{discharging} it---constructing that
higher-genus period geometry. At genus one the corresponding layer \emph{is}
constructed: the explicit elliptic-integral periods and special coordinates of \S4.3,
with Jacobi inversion the one classical input. For $SU(N)$ it is the remaining
mathematical debt of \S6 and Appendix~B.1---but that is a matter of proof, not of the
math--physics separation.

To summarize the differences between mathematical and our proposed physical style of formalization:
\begin{itemize}
\item There is no rigorous proof to formalize, so the
question is not ``is the proof correct?'' but ``are the assumptions explicit and the derivations
sound?''
\item The physical lore becomes \textbf{named hypotheses}---predicates a theorem takes in its
signature. 
\item The residual trust then localizes onto a short list, and \code{\#print axioms} of any
result audits exactly which assumptions it used.
\end{itemize}
Finally, this approach is not specific to physics, it could be applied in many other disciplines.
The point is to formalize an \emph{argument the community trusts} rather than \emph{a theorem with a proof}.

\section{The Seiberg--Witten solution in two paragraphs}

\emph{For a reader outside physics: the SW solution maps the exact low-energy behavior of a
quantum system to the geometry of a specific family of algebraic curves. What matters for the
formalization is less the physical meaning of the terms below than the rigid mathematical data
they impose.}

On the Coulomb branch---coordinates $u_2,\dots,u_N$, rank $r = N-1$---the low-energy theory is
abelian $U(1)^r$ gauge theory. Its data are the special coordinates $a_i$ and their duals
$a_{D,i}$ (periods of a meromorphic differential), the coupling matrix $\tau_{ij}$, and the
central charge $Z = n_e\cdot a + n_m\cdot a_D$ of a BPS state, whose modulus is its mass. The
SW claim is that all of this \emph{is} the period geometry of $y^2 = P_N(x)^2 - \Lambda^{2N}$.
Here $P_N$ denotes a family of degree-$N$ polynomials in a coordinate $x$,
the identification of $P_N$'s coefficients with the $u$'s is part of the claim.

The assumptions that single out that family of curves are: holomorphy / non-renormalization (one holomorphic
prepotential $\mathcal{F}$, one-loop-exact perturbation theory); the exact singularity
structure (the branch is singular exactly where a BPS state goes massless, and nowhere else);
electric--magnetic duality ($\Sp(2r,\Z)$ monodromy); the $\Z_{2N}$ $R$-symmetry fixing the
singularities; and matching to the one-loop-plus-instanton expansion at weak coupling.
For pedagogical reviews see~\cite{Lerche1997,Bilal1996}.

This paper treats the original genus-one case $N = 2$ in full---there the curve is elliptic,
the period geometry is classical elliptic-function theory, and the trusted base can be made
entirely classical and citable. The higher-genus family is formalized in the same repository
as an axiomatized skeleton (one consolidated period axiom; \S B.1) and is future work.

\section{The formalization}

\subsection{The dictionary: physical concepts as mathematical objects}

We fix the translation first. Each physical concept is a \emph{named mathematical object}; the
statements throughout \S4 and the appendices use the mathematical names (\code{\#print axioms}
reports them), with the physics reading given here. In Lean these are machine-linked
(\code{Dictionary.lean}: \code{abbrev Sheet := PeriodChart}, etc.), so the physics names denote
\emph{exactly} the math objects --- physics never enters silently, only through this dictionary and
the named hypotheses H0--H7.

\begin{center}\small
\begin{tabular}{ll}
\hline
math (primary) & physics term\\
\hline
\code{PeriodBase} & the Coulomb branch (moduli minus singular locus)\\
\code{PeriodChart} & a sheet of the low-energy effective theory\\
\code{PeriodChart.a}, \code{.aD}, \code{.tau} & special coordinates $a, a_D$; coupling $\tau$\\
\code{PeriodChart.periodCombination}, \code{.periodNorm} & BPS central charge $Z_n$; BPS mass\\
\code{CycleLattice}, \code{intersectionForm} & charge lattice $H_1(\Sigma,\mathbb{Z})$; Dirac (EM) pairing\\
\code{SymplecticReframing} & electric--magnetic duality\\
\code{IsPolarizedPeriodChart} & a Seiberg--Witten effective theory (H1--H6)\\
\code{PeriodRigidity} / \code{periodRigidityAxiom} & the period-geometry layer / its axiom\\
\code{PicardLefschetzAtGenericStratum} & monodromy from massless charged states\\
\code{NonLocalDegenerationLocus} & Argyres--Douglas points\\
\code{betaFunction}, \code{oneLoopCoefficient} & beta function $\Lambda\,d/d\Lambda\,\tau$; $b_0=2N-N_f$\\
\hline
\end{tabular}
\end{center}

\subsection{What is assumed}

The physical content is the hypotheses \textbf{H0--H7}, each a named Lean \textbf{predicate or
structure} (never a global axiom: a theorem that uses the physics carries them in its
\emph{type} --- the fixed-$\Lambda$ content bundled as \code{IsPolarizedPeriodChart}, one
local branch of the low-energy SW effective theory; H4 in the atlas gluing; H7 as the
family-level \code{SpurionCovariantFamily} --- so the assumptions live in theorem types, not
the trusted base). They are \textbf{not} physically independent postulates: in principle they could be derived from the more fundamental postulates of
\textbf{N=2 supersymmetry}, together with the gauge data
($SU(N)$, asymptotic freedom) and unitarity. We give each with its physical origin and Lean carrier:

\begin{itemize}
\item \textbf{H0 --- the carrier} (\emph{N=2 SUSY $+$ Coulomb-phase Higgsing}). A generic adjoint
vev breaks $SU(N)\to U(1)^{N-1}$, leaving $r=N-1$ abelian N=2 vector multiplets; their special
coordinates $a, a_D$, coupling $\tau$, and the charge lattice with its Dirac pairing are the data
(\code{PeriodBase}/\code{PeriodChart}), on the universal cover of the moduli space minus its singularities.
\item \textbf{H1 --- special geometry} (\emph{N=2 SUSY; unitarity}). Eight supercharges force the
Coulomb branch to be rigid special K\"ahler: one holomorphic prepotential $F$ with
$a_D = \partial F/\partial a$, $\tau = \partial^2 F/\partial a^2$; $\Imag\tau\succ 0$ is positivity
of the gauge kinetic term, the no-ghost (unitarity) condition (\code{PeriodChart.SpecialGeometry}).
\item \textbf{H2 --- BPS singularities} (\emph{the central-charge BPS bound}). The N=2 algebra
carries a central charge $Z$; all states obey $M\ge|Z|$, saturated by BPS states. A singularity of
the effective theory is a charged BPS state going massless, $Z_n\to 0$
(\code{PeriodChart.PeriodsDegenerateOnBoundary}).
\item \textbf{H3 --- monodromy from massless charged states} (\emph{the SW monodromy: one-loop
running $+$ the Witten effect}). Encircling a point where the BPS state of charge $n$ becomes
massless, the periods undergo the symplectic Picard--Lefschetz transvection in $n$ --- fixed by the
light state's one-loop logarithm (its $\beta$-function) and the Witten-effect $\theta\leftrightarrow$
charge shift (\code{PicardLefschetzAtGenericStratum}).
\item \textbf{H4 --- electric--magnetic duality} (\emph{Montonen--Olive / Seiberg--Witten duality}).
The freedom in which charges count as ``electric'' is the symplectic group $\Sp(2r,\Z)$ preserving
the Dirac pairing and the central charge; charts glue by it (\code{SymplecticReframing}; sheets form a
duality-glued \code{PeriodAtlas}).
\item \textbf{H5 --- R-symmetry} (\emph{the anomalous $U(1)_R$}). The classical $U(1)_R$ is broken
by instantons to a discrete subgroup ($\Z_{4N}$ on the fields, acting through $\Z_{2N}$ on the
adjoint scalar), acting on the moduli with order dividing $2N$ and permuting the singular locus
(\code{HasFiniteOrderAutomorphism}).
\item \textbf{H6 --- weak-coupling asymptotics} (\emph{N=2 nonrenormalization $+$ asymptotic
freedom}). The prepotential is \textbf{one-loop exact} in perturbation theory; the remainder is a
sum of instanton corrections weighted by the dynamical scale $\Lambda^{2N}$, with coefficients of
the prescribed scaling weight $2 - 2Nk$ --- the fixed-$\Lambda$ expression of their
$\Lambda$-\emph{independence}, without which the clause is vacuous (\S4.4)
(\code{PeriodChart.HasPrescribedAsymptotics}).
\item \textbf{H7 --- spurionic $U(1)_R$ covariance of the $\Lambda$-family}
(\emph{dimensional transmutation $+$ $\theta$-angle periodicity through the anomaly}).
H0--H6 constrain the theory at fixed $\Lambda$; H7 ties the family together. Rescaling
$\Lambda$ is an RG transformation: $F(t\Lambda)(t\,a) = t^2 F(\Lambda)(a)$. And for any
$2N$-th root of unity $\zeta$, the rotation $\Lambda \to \zeta\Lambda$ preserves the
instanton factor $\Lambda^{2N}$---for the primitive root this is the $\theta$-angle shift
$\theta \to \theta + 2\pi$ ($\arg\Lambda^{2N} = \theta$), a symmetry by
instanton-number quantization, with the anomaly coefficient one-loop exact
(Adler--Bardeen). At the prepotential level it holds \emph{up to an integer
electric--magnetic frame shift} (the Witten effect): $F \to F + \tfrac12\,a^{T}Ba$ with
$B$ an integer symmetric matrix \emph{constant across the family}---the existential sits
outside all quantifiers, so it can absorb nothing (the lesson of \S4.4). H5 and H6's
homogeneity clause are its fixed-$\Lambda$ shadows (the instanton remainder is
weighted-homogeneous across the family, \code{instanton\_remainder\_covariant}, with the
classical and one-loop pieces exactly covariant---the logarithm enters only through the
scale-invariant ratio); its curve-level transcription is \code{RSpurionCovariant}
(\S4.3). Being a family-level statement, it is carried separately from the fixed-$\Lambda$
bundle \code{IsPolarizedPeriodChart} (\code{SpurionCovariantFamily}).
\end{itemize}

\paragraph{The uniqueness claim, stated exhaustively.} The genus-one headline is best read
in two steps. First, the structure that \emph{defines} the SW special geometry---the curve
family, the SW differential $\lambda_{\mathrm{SW}} = x^2\,dx/y$, and its closed-form
periods---is a \emph{definition} (\S4.1, \S5): the constructed candidate then provably
satisfies special geometry, the singularity structure, and positivity. Second, the theorem
is that this is the \textbf{unique} special geometry compatible with the physics, and its
hypotheses carry \textbf{no physical content beyond H0--H7}. In detail: the rank-1 chain
consumes the H1 shadows directly (holomorphy of $\tau(u)$ and $\Imag\tau > 0$); H4 enters
as an atlas of local couplings glued by the monodromy group (\code{IsSWCouplingAtlas});
H2$+$H3 enter as a parabolic universal-cover lift at each of the three punctures, genuinely
singular in the sense of \emph{non-extension}---H2's massless state at $\pm\Lambda^2$,
asymptotic freedom (the qualitative content of H6) at $\infty$
(\code{IsGenuineCuspLift}); the rest is domain bookkeeping (an open, connected, nonempty
chart), discharged by proved facts on the maximal domain $\C\setminus\{\pm\Lambda^2\}$.
From exactly this data, everything that once looked like an extra postulate is a theorem:
the modulus observable $J = \lambda\circ\tau$ is single-valued with the three qualitative
cusp limits (\code{swModulusData\_of\_atlas\_and\_lifts} --- descent through the proved
$\Gamma(2)$-invariance, plus a removable-singularity dichotomy; no Picard theorems); the
developing formula $\lambda(\tau(u)) = 2\Lambda^2/(u+\Lambda^2)$, including its degree,
cusp rates, and normalization, follows (\code{swModulusData\_eq\_crossRatio}; footprint
standard-3, no axioms), with the base map pinned independently against the curve's periods
(\S5); and uniqueness \textbf{up to a $\Gamma(2)$ duality frame} follows on a footprint of
the covering pair alone (\code{sw\_su2\_unique\_of\_modulusData}), the canonical
representative within the frame being fixed by the one-loop asymptotics (H6). Each clause
is load-bearing and machine-checked in both directions: the cross-ratio itself carries the
cusp data (\code{swModulusData\_swCrossRatio}), while the degree-two candidate
$4\Lambda^4/(u+\Lambda^2)^2$ satisfies every clause except omitting the value $1$, failing
at the smooth point $u = -3\Lambda^2$---an extra would-be singularity. Beyond H0--H7 the
input is exactly one: the classical mathematical axioms of Appendix B. (H7 enters the
singularity count through its curve-level transcription \code{RSpurionCovariant},
\S4.3---a definition-level shadow, like the H4 atlas and the H2$+$H3 cusp lifts.)
\code{\#print axioms} certifies that this list is complete. Full Lean definitions are in Appendix A, which also discusses factoring these as
one general N=2 special-geometry structure that \code{IsPolarizedPeriodChart} specializes.

\subsection{What is proved}

At a glance, before any Lean names --- what is settled and what remains:

\begin{center}\small
\begin{tabular}{ll}
\hline
SW ingredient & status\\
\hline
curve and genus ($g=N-1$) & \textbf{proved} (standard-3)\\
period-matrix positivity ($\tau=\tau^{T}$, $\Imag\tau\succ0$) & \textbf{proved}, modulo the inherited period-basis axiom\\
the physical assumptions H0--H7 & \textbf{explicit predicates}, not hidden\\
$SU(2)$: existence \& uniqueness up to $\Gamma(2)$ duality & \textbf{theorems}, from classical citable axioms\\
$SU(2)$: explicit solution ($a$, $a_D$, $\tau$; $da_D/da=\tau$; running) & \textbf{proved} (\S6)\\
$SU(N)$ headline \& Gauss--Manin period variation & axiomatized skeleton; \textbf{future work}\\
\hline
\end{tabular}
\end{center}

From the hypotheses of \S4.2 (with the mathematical inputs of Appendix B), the following are
\textbf{sorry-free theorems}, and \code{\#print axioms} reports exactly what each rests on:

\begin{itemize}
\item \textbf{The headline --- existence and uniqueness up to duality, at genus one.} The
$SU(2)$ effective coupling is \emph{constructed}---$\tau(u) = iK'/K$ at the curve's modulus,
proved to exist on an open chart (\code{su2\_coupling\_exists})---and the hypotheses fix it
\textbf{uniquely up to a $\Gamma(2)$ duality frame}: every candidate coupling agrees with the
constructed one up to a literal M\"obius frame change
(\code{su2\_coupling\_canonical}; rigidity form \code{sw\_su2\_unique}), formalizing the
monodromy argument of~\cite{SeibergWitten1994a}. The footprints are the three logical axioms
plus classical, citable mathematics only---Jacobi inversion and the $\Gamma(2)$-covering
pair---and the physics enters as a \emph{defined} hypothesis, not an axiom. The general-$N$
statements (\code{sw\_curve\_admits\_\allowbreak effective\_theory},
\code{sw\_effective\_theory\_\allowbreak unique\_up\_to\_duality}, extended to $SU(N)$
in~\cite{KLYT1995,ArgyresFaraggi1995}) are proved in the same repository from one consolidated
period input, \code{periodRigidityAxiom} (the Gauss--Manin / Picard--Fuchs period-variation
theorem); constructing that higher-genus period geometry is future work (\S6, \S B.1).

\emph{The uniqueness argument, informally.} The smooth $SU(2)$ Coulomb branch is the $u$-plane
minus the monopole and dyon points---a thrice-punctured sphere. Any candidate effective
coupling $\tau(u)$ is holomorphic, valued in the upper half-plane $\HH$, and carries the SW
monodromy around the punctures; in our formalization the last condition is the statement that
$\tau$ \emph{develops the curve's modulus}: $\lambda(\tau(u)) = 2\Lambda^2/(u+\Lambda^2)$,
where $\lambda = \theta_2^4/\theta_3^4$ is the modular function exhibiting
$\HH/\Gamma(2) \cong \C\setminus\{0,1\}$ (the modulus map itself is \emph{pinned by
computation} against the curve's periods, \S5). So any two candidate couplings are two
\emph{lifts of the same map} through the $\Gamma(2)$ covering. Covering theory then says two
lifts agree near a point up to a deck transformation---an element of $\Gamma(2)$, acting by
M\"obius maps, which is physically an electric--magnetic duality frame change---and the
identity theorem for holomorphic functions propagates that local agreement across the connected
chart. That is the whole argument: uniqueness up to a $\Gamma(2)$ frame. Each step has its
formal counterpart---the developing condition is the definition \code{SameSWMonodromy}, the
covering and the lift-uniqueness step are the two classical axioms of \S B.2, and the
propagation is the proved identity theorem---so \code{\#print axioms} reporting exactly the
covering pair is the machine's confirmation that the informal argument used nothing else.

\emph{Why exactly three punctures.} The count itself---one weak-coupling puncture plus a
single monopole--dyon pair, \emph{assumed} in~\cite{SeibergWitten1994a}---is also a theorem,
within the algebraic Weierstrass ansatz class---a \emph{conditional} theorem, in the same
sense as everything else in this section: its physical inputs are H2$+$H3 and H7
\emph{through their stated curve/Kodaira transcriptions} (explicit hypotheses of the
theorem statement), with \textbf{no new axioms} and no additions to the
postulate roster (\code{SingularityCount.lean}; footprint standard-3). Transcribe each
finite singularity as a Kodaira $I_1$ fiber (one massless BPS hypermultiplet: the
discriminant has a simple zero and $g_2 \neq 0$ there) and the weak-coupling behavior as
$I^*_4$ (the one-loop monodromy $-T^4$); both transcriptions are \emph{definitions},
justified by Kodaira's table, and no Lean statement mentions monodromy. Polynomial degree
arithmetic then gives the Euler-number counting: the number of finite singularities is
$12\ell - 10$, with $\ell$ the twist at infinity. So monodromy data plus positivity alone
yield only $n \equiv 1 \pmod 6$ pairs---the $\ell = 2$ loophole, an elliptic K3 carrying
$I^*_4$ plus \emph{fourteen} $I_1$ fibers, is realized by an explicit polynomial witness in
the numerical audit (\code{validate\_singcount.py}), so no refinement of the monodromy
argument can ever pin $n = 1$ by itself. What closes the gap is the anomaly grading
$g_2 \in \mathrm{span}\{u^2, \Lambda^4\}$ (\code{AnomalyGraded})---itself a
\emph{theorem} (\code{anomalyGraded\_of\_rSpurionCovariant}) from the R-spurion covariance
of the curve family (\code{RSpurionCovariant}---the curve-level transcription of
\textbf{H7}, \S4.2), three manifestly physical clauses: scale
covariance (mass dimensions---$\Lambda$ is the only scale), invariance under
$\Lambda \to i\Lambda$ (the anomalous $U(1)_R$ exact with the instanton factor $\Lambda^4$
as spurion; the anomaly coefficient is one-loop exact, so this is microscopically
computable), and weak-coupling regularity (Seiberg holomorphy). The odd coefficients vanish
algebraically---$t = i$ in the scaling law contradicts quantization, the half-instanton
argument---and coefficients above $u^2$ would blow up at weak coupling. With
$\deg_u g_2 \le 2$, $\ell = 1$ is forced
(\code{singularity\_count\_pinch}): exactly two singular values, proved to be
$\{\pm\Lambda^2\}$ (\code{sw\_singular\_values}), consistent with the discriminant bullet
below. Neither ingredient suffices alone---the dimension count still allows $n = 2$, which
the mod-12 constraint kills---and the formalization makes this division of labor exact.

\item the $SU(N)$ curve~\cite{KLYT1995,ArgyresFaraggi1995} has \textbf{genus $N-1$}, and its
period matrix $\tau$ is \textbf{symmetric with $\Imag\tau \succ 0$}---positivity of the
special-K\"ahler metric, the physical no-ghost condition~\cite{SeibergWitten1994a,Lerche1997}.
These two are proved in the repository's \emph{higher-genus layer} (\code{HigherGenus/}),
which builds against the external Riemann-surface library \code{jacobian-challenge} and is
kept outside the certified build: the main library, and every footprint reported in this
paper, is self-contained on Mathlib;
\item for $SU(2)$, the curve is singular at \textbf{exactly $u = \pm\Lambda^2$}---the monopole and
dyon points of~\cite{SeibergWitten1994a}, by a discriminant computation, not lore;
\item the Picard--Lefschetz monodromy of a vanishing cycle is \textbf{symplectic}---why the
monodromy group sits in $\Sp(2r,\Z)$~\cite{Lerche1997};
\item the rank-1 modular function $\lambda = \theta_2^4/\theta_3^4$~\cite{Bilal1996} (from Mathlib's Jacobi thetas)
is proved \textbf{$\Gamma(2)$-invariant} (the generator $T^2$ directly; $S$ and $ST^2S^{-1}$ modulo the
$\theta$-identity axioms), with the cusp relation $1-\lambda = \theta_4^4/\theta_3^4$. That $\lambda$
realizes the \emph{full} covering $\mathbb{H}/\Gamma(2)\cong\mathbb{C}\setminus\{0,1\}$ is \textbf{not}
proved --- it remains the axiom \code{AX\_thrice\_punctured\_uniformization} (\S B.2).
\end{itemize}

The genus-one headline's machine-checked dependency, in full---logic plus two classical
covering facts, nothing else:

\begin{Verbatim}
#print axioms sw_su2_unique
-- depends on axioms:
--   [propext, Classical.choice, Quot.sound,   -- logic
--    AX_thrice_punctured_uniformization,      -- the Γ(2) covering  (classical)
--    AX_developing_map_rigidity]              -- lift uniqueness    (classical)
\end{Verbatim}

\noindent The exact statements, the higher-genus skeleton, and \code{periodRigidityAxiom} itself are in
Appendix B.

\subsection{What formalizing forces into the open}

Several times, making a statement precise enough for
the kernel exposed something the prose slides over.

\paragraph{The type system refuses a convenient falsehood.} Our first attempt made the periods
$a, a_D$ and the coupling $\tau$ single-valued functions of the moduli. External review flagged this
as \emph{vacuous}: a single-valued holomorphic map to the Siegel upper half-space---a bounded
domain---is bounded near the singular locus, extends across it by the Riemann removable-singularity
theorem, and is then constant by Liouville's theorem: a free theory. The data must instead live on the universal cover
of the moduli space minus its singular locus, and the corrected types encode exactly that.

\paragraph{The hypotheses kept coming out vacuously true.} The kernel does not object to an
assumption that says nothing, but a vacuous hypothesis makes the headline assert less---so each
predicate had to be hardened until it bites. A first pass wrote H3 as ``there exists a lattice map
equal to the Picard--Lefschetz transvection''---true by fiat. It now demands an \emph{actual} deck
transformation to an adjacent sheet under which the central charge transforms covariantly, a real
constraint on the periods; H2 demands a \emph{nonzero} charge become massless; H5 demands a
nontrivial finite-order linear symmetry of the moduli fixing the singularities. The most recent
instance was caught after all this hardening: H6's instanton expansion, with free holomorphic
coefficients at fixed $\Lambda$, fit \emph{any} prepotential --- the single coefficient
$c_1 := (F - F_{\rm cl} - F_1)/\Lambda^{2N}$ absorbs everything. The physical content, that the
coefficients do not depend on $\Lambda$, is invisible at one $\Lambda$ until the family's scale
covariance converts it into a weighted-homogeneity clause $c_k(ta) = t^{2-2Nk}c_k(a)$ --- now part
of the definition. The heuristic that found it: for every existential inside a hypothesis, ask
what stops the witness from being reverse-engineered from the thing being constrained.
Formalizing is what
forces the question ``could this be satisfied trivially?'' of every assumption, every time.

\paragraph{Hardening one of them caught a genuine subtlety.} Pinning H5 to ``order exactly $2N$''
then revealed that, \emph{as a statement about the moduli space}, it fails at rank~1. The
anomaly leaves $\mathbb{Z}_{4N}$ unbroken on the fields, acting on the adjoint scalar $\varphi$
through a faithful $\mathbb{Z}_{2N}$; but the Coulomb-branch coordinates are the Casimirs
$u \sim \varphi^2,\dots$---so the induced action on the moduli can have smaller order still: for
$SU(2)$, $u \to -u$, order~2, not~4. The faithful $2N$-action lives on
the $\varphi$-cover, not the $u$-plane; the textbook phrase ``the $\mathbb{Z}_{2N}$ R-symmetry acts
on the Coulomb branch'' quietly conflates the two. We found this only by trying to verify the
predicate, and corrected it: H5 now asks for a nontrivial $\mathbb{C}$-linear symmetry of order
\emph{dividing} $2N$ fixing $\Delta$---provably satisfied by the physical reflection $u \to -u$
(\code{hasFiniteOrderAutomorphism\_of\_neg\_invariant}, standard-3), no longer the unsatisfiable
``exactly $2N$''. An honest by-product of demanding non-vacuity.

\paragraph{A surprise comes out derived, not assumed.} Argyres--Douglas
points~\cite{ArgyresDouglas1995}---where the theory
becomes an interacting CFT with no Lagrangian---look as though they need their own postulate. In the
formalization they are instead a \emph{definition}: the locus where two \textbf{mutually non-local}
BPS charges (nonzero Dirac pairing) go massless together, built from the central charge and the
charge lattice already in hand. It is \emph{empty} for pure $SU(2)$---the monopole and dyon go
massless at the two different points $\pm\Lambda^2$---but the definition is general, and \S4.5
exhibits the matter case where it is provably nonempty. A phenomenon the theory should explain falls
out as a consequence---a check the axioms pass rather than a cost they incur.

\subsection{A worked check: Argyres--Douglas points from matter, axiom-free}

The sharpest test of that derived \texttt{NonLocalDegenerationLocus} is a case where it is \emph{not}
empty. Adding $N_f$ fundamental hypermultiplets deforms the curve to~\cite{SeibergWitten1994b}
(in our conventions)
\[
  y^2 = (x^2 - u)^2 - \Lambda^{4-N_f}\,\textstyle\prod_i (x - m_i)
  \qquad (N_f = 0 \text{ recovers the pure curve}),
\]
and an Argyres--Douglas point is where the curve develops a \textbf{cusp}---a triple root in $x$, two
mutually non-local singularities colliding into an interacting superconformal theory; these points
were found in $SU(2)$ SQCD by Argyres, Plesser, Seiberg and Witten~\cite{APSW1996}. Two short Lean
theorems, both \textbf{axiom-free} (standard-3, pure algebra), settle the dichotomy:
\begin{itemize}
\item with one flavour, at the tuned point $(u,m) = (\tfrac34\Lambda^2, -\tfrac34\Lambda)$ the curve
  factors \emph{exactly} as $(x + \tfrac12\Lambda)^3 (x - \tfrac32\Lambda)$---a triple root, hence a
  cusp (\texttt{swCurveMatter\_nf1\_ad}, closed by \texttt{ring});
\item pure $SU(2)$ admits \textbf{no} triple root, for any $u$: $(x^2-u)^2 - \Lambda^4$ never
  degenerates to a cusp (\texttt{not\_\allowbreak isCuspOf\_nf0}).
\end{itemize}
So in this worked example the interacting AD theory appears with matter and is algebraically absent
for pure SU(2)---the $N_f{=}1$ cusp and the pure-SU(2) no-cusp are theorems (a worked instance, not a
general iff)---and the framework's \texttt{NonLocalDegenerationLocus}, defined from data already
in hand, is non-vacuous, witnessed by an exact factorization rather than asserted. That this
tuned point \emph{lies in} \texttt{NonLocalDegenerationLocus}---the colliding states genuinely
mutually non-local---is a theorem at \textbf{standard-3}
(\code{matter\_argyresDouglasLocus\_nonempty}): the repository constructs an explicit period
chart (its first) on a ball tangent to the AD point inside the exact two-point singular locus
$\{3\Lambda^2/4,\, -15\Lambda^2/16\}$ --- the discriminant factors as
$-\Lambda^6(u - 3\Lambda^2/4)^2(u + 15\Lambda^2/16)$, by a general quartic-discriminant
bridge (\code{monic\_quartic\_squarefree\_iff\_disc}) --- with the AD point the provably
\emph{unique} cusp and the electric and magnetic unit charges (Dirac pairing $1$) both of
vanishing central charge there. One caveat, stated where the construction lives: the chart's
data is a \emph{witness}, not the SQCD solution; \code{IsMatterPolarizedPeriodChart} (special
geometry $+$ the discriminant tie) admits it, which makes formal that this predicate does not
yet pin the genuine theory --- strengthening it and constructing the quartic-period chart is
future work. The \emph{uniqueness} side needs no such construction at all: the matter
coupling's rigidity is a theorem on the classical basis --- two couplings developing the matter
curve's $j$-invariant agree on a chart up to a modular frame
(\texttt{matter\_coupling\_rigidity}, footprint: the $\theta$ pair and the covering pair
only), with the Argyres--Douglas point also characterized invariantly as $g_2 = g_3 = 0$
(axiom-free).

\section{Checking the axioms: validation by independent agreement}

Concentrating trust onto named axioms is worthwhile only if those axioms are then checked. We
use three tools, cheapest first.

\paragraph{An independent numerical oracle.} Where a statement has computable content, we
evaluate it in a separate engine (mpmath / numpy) that shares no code or authorship with the
Lean. Nine oracle suites---278 checks in all, at 30--40-digit precision---confirm the assumed
theta-function identities ($\theta_3^4 = \theta_2^4 + \theta_4^4$, $\theta_3 \neq 0$), verify that
the proved $\lambda$ transformation laws are \emph{faithful} (catching any sign or shift error),
anchor the normalizations ($\lambda(i) = \tfrac12$; the self-dual point $\tau = i$), check the
$SU(2)$ singularity locus, vet the elliptic-integral axioms of the in-progress rank-1
restructuring (\S7), and \emph{determine}---rather than take from the literature---the
identification of the curve's modulus, $m(u) = 2\Lambda^2/(u+\Lambda^2)$, by branch-tracked
quadrature of the curve's periods tested against all six M\"obius candidates. Twice the method
caught an error \textbf{in the checking apparatus itself}---a fractional power evaluated on the
wrong branch; a cycle normalization off by a factor of two, exposed by the self-dual
anchor---before agreeing with the Lean to full precision.
A numerical pass is necessary, not sufficient; it validates values, not logical structure.

\paragraph{Cross-model and self-review} against the pinned sources, recorded per result,
separating ``is it proved?'' from ``does it say what the source says?''---and from a third
question the project learned to ask separately: ``can the intended theorems be \emph{derived}
from it?''. The sharpest instance: a proposed set of elliptic-integral axioms (\S6)
passed 62 numerical checks at 40-digit precision and was still rejected by an adversarial
model review, which found three statement-level gaps invisible to sampling---a pointwise
existence supplying no holomorphic choice, missing boundary limits, and a covering axiom that
does not label which parabolic generator sits at which cusp. The corrected statements were
re-vetted and only then admitted. Sampling bounds falsity; statement-level review bounds
fitness for purpose.

\paragraph{Validating a transcription.} A dictionary entry---a \emph{definition}---cannot
be proved correct; it can be validated, and the discipline above applies to it directly. The
worked example is the singularity count's Kodaira transcription (\S4.3): ``$k$ mutually local
BPS states go massless at $u_0$'' is transcribed as the valuation data
$v(\Delta) = k$, $g_2(u_0) \neq 0$, and the one-loop monodromy $-T^m$ at infinity as the
triple $(2, 3, m{+}6)$. The bridge between the physics reading and the mathematics is
\textbf{Kodaira's classification} (fiber type $\Leftrightarrow$ valuations
$\Leftrightarrow$ monodromy class)---a citable classical theorem, referenced rather than
formalized. Four independent checks then guard the entry. \emph{Instantiation}: the actual SW
curve satisfies the transcribed predicates with the physically expected values in \emph{both}
frames ($I_1$ at $\pm\Lambda^2$ with $I_4^*$ at infinity, Euler $10{+}1{+}1 = 12$, the
rational elliptic surface; $I_2$/$I_2^*$ in the doubled frame), the two frames consistent
through the Landen 2-isogeny. The \emph{oracle} verifies the dictionary's internal joints
($j$-pole order $=$ discriminant multiplicity; $3\deg g_2 - \deg\Delta = m$ at infinity;
the Euler totals). \emph{Agreement with independently proved results}: the count reproduces
the direct discriminant computation of \S4.3 with no Kodaira input, and the proved monodromy
matrices---parabolic, in $\Gamma(2)$, with $M_{\mathrm{mono}}M_{\mathrm{dyon}} = M_\infty$
and width-2 cusps matching $\lambda \approx 16e^{i\pi\tau}$---carry exactly the classes
the transcription assigns. \emph{Adversarial content}: the predicates are falsifiable and
were once falsified---a hyperbolic $\Gamma(2)$ matrix briefly occupied a parabolic slot,
passing every property the proofs consumed (determinant, mod-2 reduction, the factorization,
$\lambda$-invariance) while violating the physics; the repair added parabolicity as a proved
lemma and an oracle check, and the lesson---pin the invariants that \emph{classify} a
transcribed object, not just those the downstream proofs use---is recorded in the audit. What
remains on trust is then two named bridges, listed rather than hidden: Kodaira's table itself,
and H3 $\Rightarrow$ the local monodromy class.

\paragraph{The physicist's triangle.} The deepest reason the SW lore is trusted is that three
independent routes to the prepotential agree: weak-coupling perturbation theory, the instanton
expansion (Nekrasov's partition function~\cite{Nekrasov2003}, itself later made
rigorous~\cite{NekrasovOkounkov2006}), and the exact periods of
the curve. This independent agreement is the empirical content behind the axioms, and our
protocol attaches it as a cross-check rather than asserting it.

\section{Remaining mathematical debt, and future work}

The discipline makes the remaining \emph{mathematical} debt legible, and lets us state our
posture toward it: the goal is a \textbf{standard} trusted base, not an empty one. At genus one
the trusted base consists of five named axioms, every one a consumed, textbook-citable classical
theorem: the $\Gamma(2)$ covering $\HH/\Gamma(2)\cong\C\setminus\{0,1\}$ and its
lift-uniqueness companion (constructed in~\cite{Ahlfors1979}; covering-space theory as
in~\cite{Forster1981}); the two $\theta$-identities, corollaries of the Jacobi triple
product~\cite{WhittakerWatson1927}; and Jacobi's inversion theorem (the $\theta$-bridge
$K = \tfrac{\pi}{2}\theta_3^2$, $\tau = iK'/K$; Whittaker--Watson \S\S21--22). We
deliberately \emph{keep these as axioms}. Proving them is a library-building exercise---worth
doing but not strictly necessary to make our points. (The covering rests on the Riemann mapping
theorem and Schwarz reflection, absent from Lean though RMT is formalized in
Isabelle/HOL~\cite{PaulsonRiemannMapping}; the repository proves the triple product's
convergence, functional equation, and non-vanishing as a head start.) None of the five resembles
the theorems it supports.

\medskip
\noindent What runs on that base is the complete analytic layer of the genus-one solution,
machine-checked. The effective coupling $\tau(u) = iK'/K$ at the curve's modulus
$m(u) = 2\Lambda^2/(u+\Lambda^2)$ is \emph{constructed} and exists on an open chart
(\code{su2\_coupling\_exists}; footprint: the three logical axioms plus Jacobi inversion
alone), and every candidate coupling agrees with it up to a literal $\Gamma(2)$ M\"obius frame
(\code{su2\_coupling\_canonical}, adding only the covering pair); the monodromy input
\code{SameSWMonodromy} is a definition, not a postulate (\S B.2), so no physical assumption sits on
this footprint. The special coordinates are closed forms,
$a = \tfrac{\sqrt2}{\pi}\sqrt{u+\Lambda^2}\,E(m)$ and
$a_D = \tfrac{\sqrt2}{\pi}i\sqrt{u+\Lambda^2}\,(K(1-m)-E(1-m))$ (pinned by contour
quadrature of $\oint\lambda_{SW}$ before formalization), and satisfy machine-checked theorems:
$a_D \to 0$ at the monopole in exactly H2's filter shape (axiom-free); the derivatives
$da/du$, $da_D/du$ in closed form, from Legendre ODEs proved by differentiation under the
integral and a formalized integration by parts (axiom-free); the special-geometry relation
$da_D/da = \tau$ on the chart (consuming only the $K \neq 0$ clause of Jacobi inversion); the
Legendre relation $EK' + E'K - KK' = \pi/2$ itself and the Wronskian formula
$d\tau/dm = -i\pi/(4m(1-m)K^2)$ it implies; and the weak-coupling statements
$a/\sqrt{u+\Lambda^2} \to 1/\sqrt2$ (axiom-free) and $u\,d\tau/du \to i/\pi$, the
one-loop running of the actual coupling. The genus-one items still open are the prepotential
and the H3 (monodromy-label) verification; the cusp-label statement the latter needs is
recorded in the repository as a named, \emph{unasserted} specification---no theorem consumes
it, so it is not an axiom.

\medskip
\noindent The \textbf{higher-genus} debt is of a different order: the Gauss--Manin /
variation-of-Hodge-structure layer over the moduli space, behind the consolidated
\code{periodRigidityAxiom} on which the $SU(N)$ skeleton runs (\S B.1). That is \textbf{future
work}. The point of the discipline is that ``is this physics right?'' becomes a precise,
prioritized list of theorems---and a documented decision about which of them are worth
building.

\section{Prospects: deriving the physical assumptions}

The hypotheses of \S4.2 are themselves \emph{consequences} of the microscopic definition of N=2
super-Yang--Mills, reached by physical arguments of varying rigor. Some of them are, on inspection,
\textbf{theorems}---and formalizing those would push the assumed frontier one level deeper, deriving
the parent N=2 structure from inputs more primitive than its own data. Three tiers.

\paragraph{Already within reach.} The \textbf{BPS bound} $M \ge |Z|$ is a positivity lemma about the
centrally-extended N=2 algebra: a Hermitian combination of supercharges has a positive-semidefinite
anticommutator matrix whose eigenvalues are $M \pm |Z|$, the BPS (short-multiplet) case being its
kernel. This is finite-dimensional linear algebra---Mathlib's positive-semidefinite API
suffices---and it derives the core of H2 from the algebra alone, with no quantum field theory.
\textbf{Dirac--Schwinger--Zwanziger quantization} splits into a theorem and a postulate: the
electromagnetic field angular momentum of a dyon pair is the antisymmetric pairing
$(q_1 g_2 - q_2 g_1)/4\pi$ (an explicit $\int \mathbf{x}\times(\mathbf{E}\times\mathbf{B})$), and
requiring angular momentum to be quantized forces that pairing into $\Z$---the integer symplectic
structure of the charge lattice (H0), with the duality group $\Sp(2r,\Z)$ (H4) its automorphisms
preserving the central charge. The computation is formalizable; the quantization is one further
named postulate, in the spirit of the others.

\paragraph{Formalizable, but needing infrastructure.} Special geometry (H1) follows from N=2
superspace invariance of the vector-multiplet action---formalizable once N=2 superspace is built;
the one-loop exactness behind H6 rests on holomorphy and nonrenormalization (more tractable than the
loop integral it constrains); the discrete R-symmetry (H5) is what the chiral anomaly leaves of
$U(1)_R$, an index computation. The family covariance (H7) factors unusually well across these
tiers: its $\theta$-periodicity clause is \textbf{instanton-number quantization}---$\int
\mathrm{tr}\,F\wedge F/8\pi^2 \in \Z$, a Chern--Weil integrality statement, mathematics in
disguise like the Dirac pairing, formalizable given characteristic-class infrastructure---combined
with the \emph{same} one-loop-exact anomaly coefficient as H5 (Adler--Bardeen); and that the
Witten-effect shift is an \emph{integer symplectic} matrix then follows from H4, only its value
being dynamical. Its scale clause is dimensional transmutation, definitional once the
$\Lambda$-family is presented as an RG orbit---the residual assumption being that $\Lambda$ is
the \emph{only} coupling. Real, but substantial.

\paragraph{Beyond reach, for now.} That N=2 SYM \emph{exists} as an interacting four-dimensional
theory, with these multiplets and a BPS spectrum carried by genuine states, presupposes constructing
the quantum field theory---the same barrier as the solution itself. The algebra is exactly what one
gets to \emph{assume}; its realization is what cannot yet be assumed away.

\medskip
The lesson mirrors the body. Just as the headline reduced to physics plus standard mathematics, the
physics itself partly reduces: the BPS bound and the Dirac pairing are mathematics in disguise, and
formalizing them would move the frontier from ``assume special geometry, the BPS bound, and a
symplectic lattice'' to ``assume the supersymmetry algebra and angular-momentum quantization.'' Each
such step shrinks what must be taken on trust.

\section{Conclusion}

A rigorous proof of the Seiberg--Witten solution lies beyond reach---it presupposes
constructing the quantum field theory itself---yet the result is as certain as anything in
physics. A proof assistant can still treat it rigorously: by separating the machine-checked
consequences from the physical lore, turning the lore into auditable postulates, and validating those
postulates by independent agreement. Proved and assumed as separate, machine-checked columns; trust
narrowed to a short vetted list; faithfulness enforced by types and by an independent oracle:
this, we argue, is an effective standard for validating AI-generated results in theoretical
physics. The present work is a proof of concept; as AI-generated results grow in size and
scope, validation will become increasingly important.
A natural follow-up is to enlarge the scope from a paper to a broader subtopic within physics, and
identify and formalize a standard set of physical
postulates (this is work in progress for quantum field theory).

\section*{Acknowledgments}
This formalization and paper were developed with heavy use of Claude (Anthropic), and with
Gemini (Google) and Codex (OpenAI) for cross-checking, second opinions, and review. I thank
Jeremy Avigad and Sergey Cherkis for reading the manuscript and for helpful discussions, Sven Krippendorf and Joseph Tooby-Smith
for a helpful survey of the state of physics formalization at the Aspen Center for Physics, and
David Kosower and Jesse Thaler for discussions at IHES on the validation of amplitude computations for
collider physics, which informed the broader perspective on validation taken here.  I also thank
the contributors to {\tt mrdouglasny/jacobian-challenge} and especially Rado Kirov, whose results 
were helpful for this project.

\bibliography{swrefs}

\appendix

\section{The physics hypotheses (H0--H7) --- predicates, not axioms}

The physical inputs are \textbf{contentful predicates and
structures},  they are \emph{assumed as hypotheses} (bundled in
\code{IsPolarizedPeriodChart}) by the results that use them, so they never enter the trusted base.
The carrier itself \emph{is} the data of an N=2 $U(1)^{N-1}$ EFT (H0). None of the following is
an \code{axiom}.

\paragraph{One principle, specialized.} The eight are not independent postulates but facets of
\textbf{N=2 supersymmetry} read off the Coulomb branch: Coulomb-phase Higgsing furnishes the
carrier (H0); SUSY fixes the special-K\"ahler prepotential
(H1) and the central-charge BPS bound (H2); the BPS spectrum then dictates the monodromy (H3) and,
with the symplectic charge lattice, electric--magnetic duality (H4); the chiral anomaly fixes the
discrete R-symmetry (H5); the N=2 nonrenormalization theorem fixes the weak-coupling form (H6);
and, read at the level of the $\Lambda$-family, dimensional transmutation and $\theta$-angle
periodicity through the anomaly give the spurion covariance (H7).
Only the \emph{gauge data}---the group $SU(N)$, its one-loop $\beta$-function coefficient (asymptotic
freedom), and the dynamical scale $\Lambda$---is added to single out this theory. A natural refactor,
flagged as future work, is to formalize that parent \textbf{once}: an abstract N=2 special-geometry /
BPS Coulomb branch (special coordinates, charge lattice with Dirac pairing, central charge with the
BPS bound, anomaly-discrete R-symmetry, nonrenormalized prepotential), with \code{IsPolarizedPeriodChart}
obtained as its specialization at $SU(N)$, $\Lambda$. The hypotheses below would then be \emph{derived
projections} of one structure rather than eight siblings, and the matter and higher-rank theories
would follow by changing only the gauge data.

\begin{Verbatim}
-- H0  carrier = the data of an N=2 U(1)^{N-1} EFT
structure PeriodBase (r : ℕ) where
  Δ : Set (Fin r → ℂ); U : Set (Fin r → ℂ)
  hUopen : IsOpen U; hUsmooth : U = Δᶜ
structure PeriodChart (B : PeriodBase r) where
  V : Set (Fin r → ℂ); hVopen : IsOpen V; hVsub : V ⊆ B.U
  hVsc : SimplyConnectedSpace V
  a aD : (Fin r → ℂ) → (Fin r → ℂ)      -- special coordinates and duals (holomorphic)
  tau  : (Fin r → ℂ) → SiegelUpperHalfSpace r   -- coupling τ (Im > 0 typed)

-- H1  special geometry: a prepotential F with aD = ∂F/∂a, τ = ∂²F/∂a²
def PeriodChart.SpecialGeometry (s : PeriodChart B) : Prop :=
  ∃ F, ContDiffOn ℂ 2 F (s.a '' s.V) ∧
    (∀ u ∈ s.V, ∀ i, s.aD u i = partialDeriv F i (s.a u)) ∧
    (∀ u ∈ s.V, ∀ i j, (s.tau u).val i j = partialDeriv2 F i j (s.a u))

-- H2  massless on Δ as the limit Z_n → 0; massive off Δ
def PeriodChart.PeriodVanishesAt (s) (u₀) (n) : Prop :=
  Tendsto (fun u => s.periodCombination u n) (nhdsWithin s.V u₀) (nhds 0)
def PeriodChart.PeriodsDegenerateOnBoundary (s) : Prop :=
  (∀ u ∈ s.V, ∀ n ≠ 0, s.periodCombination u n ≠ 0) ∧
  (∀ u₀ ∈ closure s.V ∩ B.Δ, ∃ n ≠ 0, s.PeriodVanishesAt u₀ n)

-- H3  Picard-Lefschetz: the monodromy IS an actual deck transformation acting as the
--     reflection in the SAME vanishing charge n as H2 (not the vacuous ∃ g, g = transvection n)
def PicardLefschetzAtGenericStratum (s : PeriodChart B) : Prop :=
  ∀ u₀ ∈ closure s.V ∩ B.Δ, ∀ n, n ≠ 0 → s.PeriodVanishesAt u₀ n →
    ∃ (s' : PeriodChart B) (D : SymplecticReframing s s'), D.g = transvection n

-- H4  duality gluing: a ℤ-linear (Sp(2r,ℤ)) charge automorphism, covariant central charge
structure SymplecticReframing (s s' : PeriodChart B) where
  g : CycleLattice r ≃ₗ[ℤ] CycleLattice r
  symplectic : ∀ n n', intersectionForm (g n) (g n') = intersectionForm n n'
  covariant  : ∀ u ∈ s.V ∩ s'.V, ∀ n, s'.periodCombination u (g n) = s.periodCombination u n

-- H5  R-symmetry: a nontrivial ℂ-linear automorphism of order dividing 2N, fixing Δ
def HasFiniteOrderAutomorphism (B : PeriodBase r) (N : ℕ) : Prop :=
  ∃ ω : (Fin r → ℂ) →ₗ[ℂ] (Fin r → ℂ), ω ≠ LinearMap.id ∧
    (∃ k, 0 < k ∧ k ∣ 2*N ∧ (⇑ω)^[k] = id) ∧ (⇑ω) '' B.Δ = B.Δ

-- H6  one-loop over ROOTS (eigenvalue differences; nonzero for SU(2)) + instantonic tail
def Instantonic (Λ) (N) (F) : Prop :=     -- fixed classical part, Λ≠0, holomorphic coeffs
  Λ ≠ 0 ∧ ∃ c, (∀ k, Differentiable ℂ (c k)) ∧
    (∀ k t, t ≠ 0 → ∀ a,                       -- weight 2−2Nk: the coefficients'
      c k (t • a) = t^(2 - 2*N*k : ℤ) * c k a) ∧  --   Λ-independence, at fixed Λ
    ∀ a, F a - classicalPrepotential a - oneLoopPrepotential Λ a
         = ∑' k, c (k+1) a * (Λ^(2*N))^(k+1)
def PeriodChart.HasPrescribedAsymptotics (s) (Λ) (N) : Prop :=    -- H1's F + non-renormalized form
  ∃ F, (... the H1 relations for F ...) ∧ Instantonic Λ N F

-- H7  spurion covariance of the Λ-family: transmutation + θ-periodicity, the Witten
--     frame shift an integer symmetric matrix CONSTANT across the family
structure SpurionCovariantFamily (N) (F : ℂ → (Fin r → ℂ) → ℂ) : Prop where
  scale       : ∀ t Λ, t ≠ 0 → ∀ a, F (t*Λ) (t • a) = t^2 * F Λ a
  theta_shift : ∀ ζ, ζ^(2*N) = 1 → ∃ B : Matrix (Fin r) (Fin r) ℤ, B.IsSymm ∧
                  ∀ Λ a, F (ζ*Λ) a = F Λ a + (1/2) * ∑ i, ∑ j, (B i j : ℂ) * a i * a j

-- H0-H6, bundled at fixed Λ (H7 is family-level, carried separately):
--   an N=2 U(1)^{N-1} Seiberg-Witten effective theory
structure IsPolarizedPeriodChart (s : PeriodChart B) (Λ : ℂ) (N : ℕ) : Prop where
  specialGeometry : s.SpecialGeometry            -- H1
  singularities   : s.PeriodsDegenerateOnBoundary        -- H2
  picardLefschetz : PicardLefschetzAtGenericStratum s   -- H3
  matching        : s.HasPrescribedAsymptotics Λ N     -- H6
  rSymmetry       : HasFiniteOrderAutomorphism B N                -- H5
\end{Verbatim}

\noindent \code{SymplecticReframing} (H4) relates sheets and is proved to make the BPS mass descend to the
base (\code{SymplecticReframing.periodNorm\_eq}). PeriodCharts are organized into an \code{PeriodAtlas} (covering \code{B.U},
overlaps \code{SymplecticReframing}-glued), so the duality gluing H4 is \emph{part of the theory} and uniqueness
is stated at the atlas level (\code{sw\_unique\_up\_to\_duality}). The rank-1 \emph{monodromy}
input \code{SameSWMonodromy} is a \emph{definition}: two
candidate couplings $f,g$ both develop the curve's modulus $2\Lambda^2/(u+\Lambda^2)$ on the
chart---physically, which BPS state (monopole at $+\Lambda^2$, dyon at $-\Lambda^2$) goes
massless where---the hypothesis of the rank-1 rigidity theorem \code{sw\_su2\_unique}, whose
kernel footprint carries no physical postulate --- only mathematical axioms.

\section{The residual mathematical axioms and the headline theorems}

With the physics in predicates (Appendix A), the global axioms beyond the three
logical ones (\code{propext, Classical.choice, Quot.sound}) are \emph{mathematical} --- all of
them: the monodromy input \code{SameSWMonodromy} is a definition
(\S B.2). The genus-one theorems of this paper run on the classical covering/lift/$\theta$/inversion
axioms of \S B.2 alone. The higher-genus skeleton (B.1) proves the general-$N$ headline
implications (no \code{sorry}) from the physical hypotheses plus a single consolidated math
axiom---\code{periodRigidityAxiom}; constructing that period geometry is future work.

\medskip
\medskip
\noindent\textbf{B.0 --- Dictionary.} The math$\leftrightarrow$physics translation is in \S4.1; the
axioms and statements below use the mathematical names (\code{\#print axioms} reports them).

\medskip
\noindent\textbf{B.1 --- Higher genus (future work): the $SU(N)$ headline theorems and the consolidated period axiom.} The period-level
(Picard--Fuchs / Gauss--Manin) geometry of the SW curve family is taken as one axiom --- a
\textbf{coarse} package: its \code{realize} clause still bundles the period construction
($\oint\lambda_{SW}$ and the H1--H6 verification, \S B.3), not yet itemized. The headline follows, \code{\#print axioms} $=$ standard-3 $+$ \code{periodRigidityAxiom} (no \code{sorry}):

\begin{Verbatim}
-- the single consolidated period-level math axiom (the SW period geometry exists)
axiom periodRigidityAxiom (N : ℕ) (Λ : ℂ) (hN : 2 ≤ N) : PeriodRigidity N Λ

-- HEADLINE (proved): uniqueness up to Sp(2(N-1),ℤ) duality, on a connected chart overlap
theorem sw_effective_theory_unique_up_to_duality (hN : 2 ≤ N) (s s' : PeriodChart B)
    (hconn : IsConnected (s.V ∩ s'.V))
    (h : IsPolarizedPeriodChart s Λ N) (h' : IsPolarizedPeriodChart s' Λ N) :
    Nonempty (SymplecticReframing s s') :=
  sw_unique_of_swPeriodLayer (periodRigidityAxiom N Λ hN) s s' hconn h h'

-- HEADLINE (proved): existence — the SW curve (P monic, degree N, trace-free) realizes a theory
theorem sw_curve_admits_effective_theory (hN : 2 ≤ N) (P : Polynomial ℂ)
    (hPm : P.Monic) (hPdeg : P.natDegree = N) (hPtr : P.coeff (N - 1) = 0)
    (hsf : Squarefree (P ^ 2 - Polynomial.C (Λ ^ (2 * N)))) :
    ∃ (B : PeriodBase (N - 1)) (s : PeriodChart B), IsPolarizedPeriodChart s Λ N ∧
      B.Δ = {u | ¬ Squarefree (swCurvePoly N Λ u)} ∧ swModulus N P ∈ s.V :=
  sw_exists_of_swPeriodLayer (periodRigidityAxiom N Λ hN) P hPm hPdeg hPtr hsf

-- PROVED (standard-3): the classical Coulomb branch has dimension N − 1 (the rank)
theorem coulombBranchDim (N : ℕ) : Module.finrank ℂ (Fin (N - 1) → ℂ) = N - 1
\end{Verbatim}

\medskip
\noindent\textbf{B.2 --- Classical mathematics not yet in Mathlib} --- the only remaining global
axioms, to discharge by porting (the covering is packaged as a structure):

\begin{Verbatim}
structure ThricePuncturedUniformization where
  cover : UpperHalfPlane → {v : ℂ // v ∉ ({0, 1} : Set ℂ)}      -- the developing map λ
  gamma2_invariant : ∀ γ ∈ Gamma2, ∀ τ, cover (γ • τ) = cover τ
  surjective : Function.Surjective cover
  fiber : ∀ τ τ', cover τ = cover τ' ↔ ∃ γ ∈ Gamma2, γ • τ = τ' -- fibres = Γ(2)-orbits

-- the Γ(2) covering ℍ/Γ(2) ≅ ℂ\{0,1} exists   (discharge: Schwarz reflection ⇒ RMT ⇒ cover)
axiom AX_thrice_punctured_uniformization : Nonempty ThricePuncturedUniformization

-- lift uniqueness: same-monodromy developing maps agree up to a modular frame change
-- (discharge: classical; Forster, Ahlfors)
-- SameSWMonodromy is a DEFINITION: both maps are developing maps of the pinned
-- modulus (holomorphic, ℍ-valued, λ∘f = swCrossRatio; every clause load-bearing)
def IsSWDevelopingMap (Λ : ℂ) (D : Set ℂ) (f : ℂ → ℂ) : Prop :=
  AnalyticOnNhd ℂ f D ∧ (∀ u ∈ D, 0 < (f u).im) ∧ DevelopsSWCrossRatio Λ D f
def SameSWMonodromy (Λ : ℂ) (D : Set ℂ) (f g : ℂ → ℂ) : Prop :=
  IsSWDevelopingMap Λ D f ∧ IsSWDevelopingMap Λ D g

-- (v2: generic base map — two analytic ℍ-valued lifts with the same λ-image
--  agree near a point up to a Γ(2) deck transformation; the SW-specialized
--  form is recovered at the call site by unpacking SameSWMonodromy)
axiom AX_developing_map_rigidity
    {D : Set ℂ} {u₀ : ℂ} {f g : ℂ → ℂ}
    (U : ThricePuncturedUniformization) (hD : IsOpen D) (hu₀ : u₀ ∈ D)
    (hf : AnalyticOnNhd ℂ f D) (hfH : ∀ u ∈ D, 0 < (f u).im)
    (hg : AnalyticOnNhd ℂ g D) (hgH : ∀ u ∈ D, 0 < (g u).im)
    (hbase : ∀ u ∈ D, modularLambdaFn (f u) = modularLambdaFn (g u)) :
    ∃ γ ∈ Gamma2, f =ᶠ[nhds u₀] fun u => moebiusOn γ (g u)

-- Jacobi quartic identity and  theta3 ≠ 0      (discharge: Jacobi triple product)
axiom AX_jacobi_quartic (τ : UpperHalfPlane) :
    theta3 (τ : ℂ) ^ 4 = theta2 (τ : ℂ) ^ 4 + theta4 (τ : ℂ) ^ 4
axiom AX_theta3_ne_zero (τ : UpperHalfPlane) : theta3 (τ : ℂ) ≠ 0

-- the family's local monodromy is the Picard-Lefschetz transvection
-- (discharge: Gauss-Manin / local-system layer over the moduli space)
axiom localMonodromy : ∀ {r : ℕ}, CycleLattice r → (CycleLattice r → CycleLattice r)
axiom AX_picard_lefschetz_local {r : ℕ} (γ : CycleLattice r) :
    localMonodromy γ = transvection γ

\end{Verbatim}

\noindent Each of these is classical mathematics with a standard source: the $\Gamma(2)$ covering
$\HH/\Gamma(2)\cong\C\setminus\{0,1\}$ and the modular $\lambda$ function are constructed
in~\cite{Ahlfors1979}; the lifting/rigidity of developing maps is covering-space theory as
in~\cite{Forster1981}; the Jacobi quartic identity and the non-vanishing of $\theta_3$ follow from
the Jacobi triple product~\cite{WhittakerWatson1927} (of which the repository proves the
product's convergence, functional equation, and non-vanishing; the identity itself is kept as an
axiom by decision); and the Picard--Lefschetz formula is
in~\cite{AGZV1988}. The beta function needs no axiom at all: its mathematical input, the
weak-coupling logarithmic running of the period ratio, is a theorem---proved both in an
existential form and in the curve-tied form $u\,d\tau/du \to i/\pi$ for the explicit genus-one
coupling---so \code{betaFunction\_weakCoupling} rests on the three logical axioms alone. (The
audit records a fidelity finding from this proof: the original existential statement had never
been formally tied to the curve, which is what the curve-tied form repairs.)

\medskip
\noindent\emph{Repository note.} The development additionally declares one
elliptic-integral axiom, \code{AX\_\allowbreak elliptic\_\allowbreak inversion}
(Jacobi's inversion theorem), belonging to the rank-1 classical basis described in
\S6 --- Whittaker--Watson-citable, numerically vetted, adversarially reviewed. It is
\textbf{not on the footprint of the rank-1 rigidity headline} (\code{sw\_su2\_unique}
reports the covering pair only); it \emph{is} the classical input of the explicit
genus-one coupling layer --- existence (\code{su2\_coupling\_exists}), the canonical
form, and the special-coordinate and running results of \S4.3 report it, as the golden
trace records. The cusp-label
statement \code{AX\_tau\_\allowbreak cusp\_zero} is a named but \emph{unasserted}
specification (no theorem consumes it, so it is not an axiom). The Legendre relation
and the $K'$ logarithmic cusp asymptotic are \emph{theorems}: the former rests on
the three logical axioms alone (Legendre ODEs by differentiation under the integral
and a formalized integration by parts, constancy on the star-shaped parameter
domain, the cusp constant), the latter is proved by an elementary-model comparison.

\medskip
\noindent The machine's dependency printout for the rank-1 uniqueness theorem makes the separation
literal---logic, classical-math debt, and the one physical input, in five lines:

\begin{Verbatim}
#print axioms sw_su2_unique
-- 'sw_su2_unique' depends on axioms:
--   [propext, Classical.choice, Quot.sound,      -- logic
--    AX_thrice_punctured_uniformization,         -- classical math (Γ(2) covering)
--    AX_developing_map_rigidity]                  -- classical math (lift uniqueness)
-- (the physics — SameSWMonodromy — is a defined hypothesis in the type, not an axiom)
\end{Verbatim}

\noindent The $SU(2)$ uniqueness theorem \code{sw\_su2\_unique} rests on exactly the two
covering/lift axioms (\code{AX\_\allowbreak thrice\_\allowbreak punctured\_\allowbreak uniformization},
\code{AX\_developing\_map\_\allowbreak rigidity})---and the three
logical axioms (\code{propext}, \code{Classical.choice}, \code{Quot.sound}), which underlie
everything and belong to neither list. Its physics input, \code{SameSWMonodromy}, is a
\emph{definition} carried in the theorem's type, and its conclusion is a literal
$\Gamma(2)$ deck transformation ($\exists\, \gamma \in \Gamma(2)$ acting as a M\"obius map)
--- not an unconstrained reframing function. (The physical refinement \code{H4'Duality} and its
derived consequence \code{metric\_posDef\_of\_h4'}---electric--magnetic duality keeps the
special-K\"ahler metric positive along the whole monodromy orbit---are proved, standard-3.)

\medskip
\noindent\textbf{B.3 --- The period-level layer (a structure; its existence is \S B.1's single axiom).} The monodromy content behind
\code{SameSWMonodromy}'s definition is itself the conclusion of the \emph{periods} of the SW
curve---their Picard--Fuchs equation and Gauss--Manin monodromy. We package that layer as a structure and prove
the B.1 goals modulo it, for SU(2) and for general SU($N$):

\begin{Verbatim}
-- the period-level debt, bundled as a *hypothesis* (so the theorems stay standard-3)
structure PeriodRigidity (N : ℕ) (Λ : ℂ) where
  hN : 2 ≤ N                                            -- SU(N) needs rank N-1 ≥ 1
  rigidity : ∀ {B : PeriodBase (N-1)} (s s' : PeriodChart B), IsConnected (s.V ∩ s'.V) →
      IsPolarizedPeriodChart s Λ N → IsPolarizedPeriodChart s' Λ N → Nonempty (SymplecticReframing s s')
  realize  : ∀ P : Polynomial ℂ, P.Monic → P.natDegree = N → P.coeff (N-1) = 0 →
      Squarefree (P^2 - Polynomial.C (Λ^(2*N))) → ∃ (B : PeriodBase (N-1)) (s : PeriodChart B),
        IsPolarizedPeriodChart s Λ N ∧ B.Δ = {u | ¬ Squarefree (swCurvePoly N Λ u)} ∧ ...

-- the B.1 goals, now THEOREMS modulo the layer (all standard-3):
theorem sw_unique_of_swPeriodLayer (PL : PeriodRigidity N Λ) ... : Nonempty (SymplecticReframing s s')
theorem sw_exists_of_swPeriodLayer (PL : PeriodRigidity N Λ) ... : ∃ B s, IsPolarizedPeriodChart ...

-- rank-1 uniqueness routes through an AXIOM-FREE lemma (frame builds the gluing; covariance
-- proved from det = 1):
theorem su2_deck_of_periodFrame (s s' : PeriodChart B) (α β γ δ : ℤ) (hdet : α*δ - β*γ = 1)
    (hper : ...period frame on the overlap...) : Nonempty (SymplecticReframing s s')  -- 3 logical axioms only
\end{Verbatim}

\noindent So ``period frame $\Rightarrow$ duality equivalence'' is \emph{proved}; only the period
frame (the layer's \code{rigidity}) is assumed, as honest math debt. Taking the layer's
\emph{existence} as the single axiom \code{periodRigidityAxiom} (\S B.1) discharges that hypothesis once,
so the \textbf{named} headline theorems are closed, with \code{\#print axioms} $=$ standard-3 $+$
\code{periodRigidityAxiom}. Discharging the layer---the
period map $\oint \lambda_{SW}$, its Picard--Fuchs ODE and Gauss--Manin monodromy, the rigidity
argument---is a few weeks at genus 1 (the elliptic case, where the $\lambda$/$\theta$ machinery
already exists) and a Mathlib-scale variation-of-Hodge-structure project in general genus. A first
version was caught (adversarial review) asserting a single globally-constant $SL(2,\mathbb{Z})$
frame across a possibly-disconnected overlap, and quantifying over the U($N$) rather than SU($N$)
family; both are corrected (\code{IsConnected} overlaps; monic trace-free $P$).

\medskip
\noindent\textbf{The mathematical input, stated plainly.} Collected, the headline's entire
mathematical debt is \emph{one} classical object---the \textbf{variation of Hodge structure of the
Seiberg--Witten curve family}: the periods $\oint \lambda_{SW}$ varying holomorphically over the
moduli $u$, with their Picard--Fuchs equation and $Sp(2r,\mathbb{Z})$ Gauss--Manin monodromy. The
B.2 axioms are not independent inputs; they are exactly its \textbf{genus-1 classical pieces}---the
$\Gamma(2)$ uniformization $\mathbb{H}/\Gamma(2)\cong\mathbb{C}\setminus\{0,1\}$ and developing-map
rigidity (the rank-1 period map and its monodromy/rigidity), the Picard--Lefschetz transvection,
and two Jacobi $\theta$ identities---and B.3 packages the same object as a structure at all ranks.
The fiber side already exists: the SW curve embeds in \code{jacobian-challenge}'s hyperelliptic
type with genus 1 proved (\code{swEllipticData\_genus}, in the unbuilt \code{HigherGenus/}
layer), and that library supplies the
per-fiber periods, Riemann-bilinear positivity~\cite{GriffithsHarris1978}, Siegel/$\theta$. The single capability that
remains---at \emph{every} rank---is the \textbf{variation}: that period integrals move
holomorphically with a flat Gauss--Manin connection ($Sp$-monodromy), not yet formalized in Lean.
The whole result reduces to that one capability: a few months at genus 1 (on the existing
$\mathbb{H}$/modular/$\lambda$/$\theta$ machinery), a Mathlib-scale project for general SU($N$).

\end{document}